\documentstyle[preprint,pra,aps,cite,epsfig]{revtex}
\newcommand{\ket}[1]{\left\vert {#1} \right\rangle}
\newcommand{\bra}[1]{\left\langle {#1} \right\vert}
\newcommand{\dotp}[2]
	{\left\langle {#1}\right\vert\left. {#2} \right\rangle}
\newcommand{\dyad}[2]{
	\left\vert {#1} \right\rangle\!\!
	\left\langle {#2}\right\vert
}
\newcommand{\bfxi}{\mbox{\boldmath $\xi$}}
\title{
	Derivation of classical capacity of quantum channel for 
	discrete information source
}
\author{
	Kentaro Kato\and 
	Masao Osaki\and 
	Osamu Hirota
}
\address{
	Research Center for Quantum Communications, 
	Tamagawa University\\
	Tamagawa-Gakuen 6-1-1, Machida, Tokyo 194-8610, JAPAN\\
	e-mail: {\tt kkato@lab.tamagawa.ac.jp}
}
\begin{document}
\maketitle
\begin{abstract}
In this letter, we prove that the classical capacity of quantum 
channel for $M$ symmetric states is achieved by an 
uniform distribution on a priori probabilities. We also 
investigate non-symmetric cases such as a ternary amplitude 
shift 
keyed signal set and a $16$-ary quadrature amplitude 
modulated signal set in coherent states.



\end{abstract}
\newpage
It is well known that the conventional information theory has 
provided many fruitful applications\cite{Gal,Cov}. The quantum 
theory of communication has lead to some profound and exciting 
new insights into both physics and communication science. In 
the pioneering works, the main theme was to formulate a 
mathematical theory for communication processes which convey 
classical information by quantum states. As a result, several 
fundamental results in quantum aspect of communication and 
information theory have been brought. On the channel coding 
theorem, Stratonovich\cite{Str78} and Holevo\cite{Hol79} 
pointed out that a true capacity of quantum measurement 
channel is greater than the maximum mutual information with 
respect to detection operators and a priori probabilities of 
information symbols, which corresponds to channel capacity in 
conventional information theory denoted as ${C}_1$. In 
addition, Levitin\cite{Lev69} conjectured and 
Holevo\cite{Hol73} proved that it is bounded by the quantity, called Holevo's bound, defined by the von Neumann 
entropy for the ensemble of signal quantum states. Recently, 
Hausladen {\it et al.}\cite{Hau96} proved that the maximum value with respect to a priori probabilities of the Holevo's 
bound is the true capacity in the case of pure states 
by introducing a typical sub-space in addition to the random 
coding scheme and the square-root measurement as a decoding. 
Thus we can understand that the capacity of quantum 
measurement channel for given set 
of signal quantum states is essentially
the maximum value with respect to a priori probabilities of von 
Neumann entropy. However, a simple question arose such that 
``how to derive the maximum value with respect to a priori 
probabilities of the von Neumann entropy for the ensemble when 
quantum states are in an infinite dimensional space". The von 
Neumann entropy of the complete information source which 
consist of signal quantum states and 
their a priori probabilities has 
strong non-linearity for a priori probabilities, because it 
is given by the eigenvalues of the density operator for
the complete information source. Thus this problem is not 
trivial one. So far, we examined numerical calculation of the 
von Neumann entropy for several kinds of sets of quantum states 
\cite{Kat96,Osa98}. In this letter, we show analytical solution 
and simplification of the maximization with respect to a priori 
probability of the von Neumann entropy for some representative 
finite sets of signal quantum states 
in an infinite dimensional space.

We now consider a model in which classical information is sent 
by quantum states, and give a formulation of our problem. Let 
$\{1, 2, \dots, M\}$ be an input alphabet and let a letter 
`$i$' in the input alphabet correspond to a quantum state 
$\hat{\rho}_i$, called a letter state, which satisfy
\begin{equation}
	\hat{\rho}_i\geq 0,\quad
	{\rm Tr}\hat{\rho}_i = 1,\quad \forall i.
\end{equation}
A complete information source is defined as an ensemble of 
letter states with a priori probabilities $\{\xi_i\}$ 
($\xi_i \geq 0,\ \sum_{i=1}^{M}\xi_i$ = 1), and the statistical 
property is represented by a density operator 
\begin{equation}
	\hat{\rho}(\bfxi) = \sum_{i=1}^{M}\xi_i\hat{\rho}_i.
\end{equation}
The von Neumann entropy of $\hat{\rho}(\bfxi)$ is defined as 
\begin{eqnarray}
	S(\hat{\rho}(\bfxi))
	&=& - {\rm Tr}\hat{\rho}(\bfxi)\log\hat{\rho}(\bfxi)
	\nonumber\\
	&=& - \sum_{j}\lambda_j(\bfxi)\log\lambda_j(\bfxi),
\end{eqnarray}
where $\lambda_j(\bfxi)$ is the eigenvalue of the density 
operator $\hat{\rho}(\bfxi)$. According to the quantum channel 
coding theorem\cite{Hau96,Hol98}, when the complete information 
source is connected to the quantum measurement channel, the 
capacity is given by 
\begin{equation}
	C = \max_{\bfxi}{\mit\Delta}S(\bfxi) ,
\end{equation}
where 
\begin{equation}
	{\mit\Delta}S(\bfxi) 
	= S(\hat{\rho}(\bfxi)) 
	- \sum_{i=1}^{M}\xi_iS(\hat{\rho}_i).
\end{equation}
In general, we can carry out the maximization by the following 
two steps. (A) Finding the eigenvalues of the density operator 
as a function of a distribution on a priori probabilities. (B) 
Finding the maximum value by some optimization techniques. 
However, it is very difficult to solve analytically because of 
the following two reasons. One of them is a difficulty of 
obtaining the eigenvalues of the density operator which is 
essential to calculate the von Neumann entropy. Other is a 
nonlinearity of the function `$-p\log p$' even if one could get 
the eigenvalues. Fortunately, if the prepared states have a 
certain symmetry in the sense of Ref.\cite{Ban97}, the 
analytical solution can be obtained. In fact, we show that the 
capacity for the ensemble of $M$ symmetric states 
is achieved by 
a uniform distribution on a priori probabilities.

$M$ symmetric states are defined by
\begin{equation}
	\hat{\rho}_i 
	= \hat{V}^{i-1}\hat{\rho}_1\hat{V}^{\dagger \ i-1},
	\quad \forall i,
	\label{sym}
\end{equation}
where 
$
\hat{V}^\dagger\hat{V} =
\hat{V}\hat{V}^\dagger =
\hat{V}^M =
\hat{I}
$ and where $\hat{I}$ stands for an identity operator.
In this case, $S(\hat{\rho}_1) = S(\hat{\rho}_i)$ for all $i$, 
so that 
${\mit\Delta}S(\bfxi) = S(\hat{\rho}(\bfxi)) - S(\hat{\rho}_1)$.
Hence, the maximization problem of ${\mit\Delta}S(\bfxi)$ 
results in that of the von Neumann entropy 
$S(\hat{\rho}(\bfxi))$.

Let us represent an arbitrary distribution on a priori 
probabilities as follows:
\begin{equation}
	\bfxi = (\xi_1,\xi_2,\dots,\xi_M) \equiv \bfxi^{(1)},
\end{equation}
and its permutations 
\begin{eqnarray}
	\bfxi^{(2)} &\equiv& (\xi_2,\xi_3,\dots,\xi_1),\nonumber\\
	\bfxi^{(3)} &\equiv& (\xi_3,\xi_4,\dots,\xi_2),\nonumber\\
	&\vdots& \nonumber\\ 
	\bfxi^{(M)} &\equiv& (\xi_M,\xi_1,\dots,\xi_{M-1}).\nonumber
\end{eqnarray}
For these distributions $\bfxi^{(i)}$, the corresponding 
density operators satisfy the next relation.
\begin{equation}
	\hat{\rho}(\bfxi^{(i)}) 
	= \hat{V}^{-(i-1)}\hat{\rho}(\bfxi)\hat{V}^{\dagger -(i-1)}.
\end{equation}
Thus, each density operator is related to $\hat{\rho}(\bfxi)$ 
with the unitary operator $\hat{V}$. By basic properties of 
von Neumann entropy such as an invariance for unitary 
transformation and a concavity with respect to a priori 
distributions\cite{Per}, we have the following inequality:
\begin{equation}
	S(\hat{\rho}(\bfxi)) 
	= \frac{1}{M}\sum_{i=1}^{M}S(\hat{\rho}(\bfxi^{(i)})) 
	\leq S(\hat{\rho}(\frac{1}{M}\sum_{i=1}^{M}\bfxi^{(i)})).
	\label{eq1}
\end{equation}
The distribution of a right side of the inequality is 
\begin{equation}
	\bfxi^\prime 
	= \frac{1}{M}\sum_{i=1}^{M}\bfxi^{(i)} 
	= \left(\frac{1}{M},\frac{1}{M},\dots,\frac{1}{M}\right).
	\label{ineq}
\end{equation}
This fact means that the capacity for $M$ symmetric states 
is 
achieved by a uniform distribution on a priori probabilities, 
because the right side of Eq.(\ref{eq1}) 
is equal to or greater than the von Neumann entropy 
for arbitrary distribution 
$S(\hat{\rho}(\bfxi))$. We have thus proved.

This result is useful to derive the channel capacity.
When $\hat{\rho}_1$ is a pure state 
represented as $\hat{\rho}_1 = \dyad{\psi}{\psi}$, 
eigenvalues of the information source $\{\lambda_j\}$ can be 
obtained from the Gram matrix consisting 
of signal quantum states. 
Hence the capacity 
is given explicitly as follows.
\begin{equation}
	C = -\sum_{j=1}^{M}\lambda_j\log\lambda_j,
	\label{cofpure}
\end{equation}
where $\lambda_j$ is 
\begin{equation}
	\lambda_j 
	= \frac{1}{M}\sum_{k=1}^{M}\bra{\psi}\hat{V}^{k-1}\ket{\psi}
		\exp\left[
			-\frac{2j(k-1)\pi{\rm i}}{M}
		\right],
		\label{cofpure2}
\end{equation}
and ${\rm i} = \sqrt{-1}$. 
The result may be also applicable to the case of mixed states, 
but it depends on the signal quantum states whether 
the capacity is analytically obtained 
like as Eq.(\ref{cofpure}) and (\ref{cofpure2}).

\begin{figure}[htb]
\begin{center}
\epsfig{file=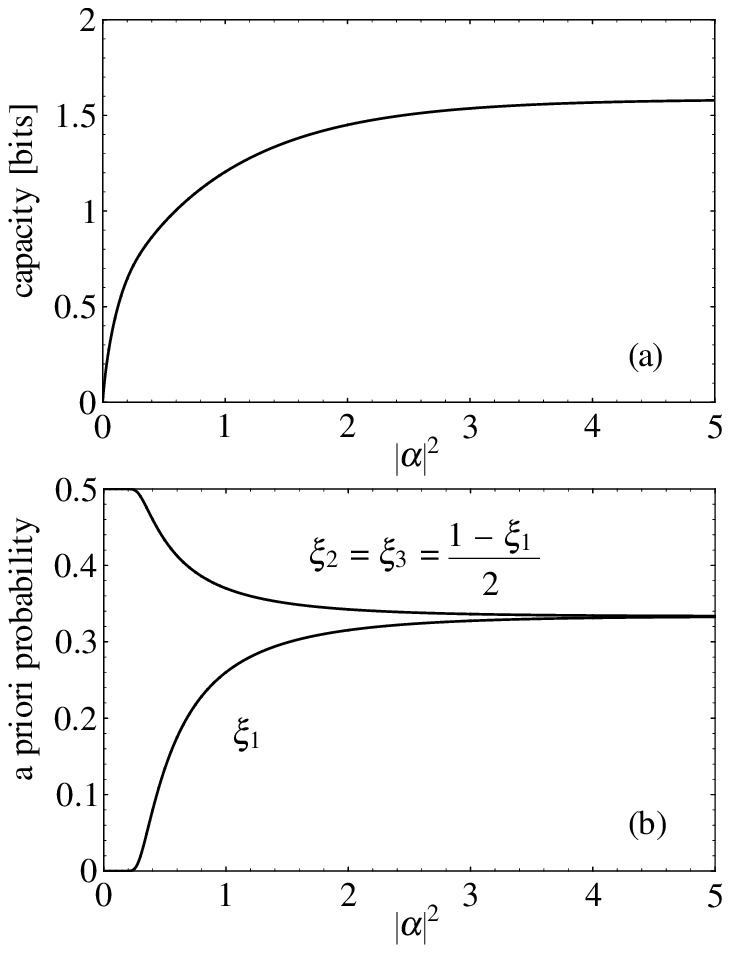,scale=1.5}
\end{center}
\caption{
Channel capacity for ternary symmetric coherent-state signal.
(a) capacity. (b) optimum distribution
}
\label{fig1}
\end{figure}

In the following, let us consider the `partially' symmetric 
case as a non-symmetric case 
to see another usage of our result. 
The word `partially' means that some states in the set of 
the prepared states have a property like as Eq.(\ref{sym}). In 
general, we cannot solve analytically in this case. However, 
the above result may provide a powerful tool for the 
calculation of the maximization. 
Here, we consider the case of a ternary 
coherent-state set, whose 
states are represented as follows. 
\begin{equation}
	\hat{\rho}_1 = \dyad{0}{0},\quad
	\hat{\rho}_2 = \dyad{\alpha}{\alpha},\quad
	\hat{\rho}_3 = \dyad{-\alpha}{-\alpha} 
	= \hat{V}\dyad{\alpha}{\alpha}\hat{V}^\dagger, 
\end{equation}
where 
$
\hat{V} = \exp[-\pi{\rm i}\hat{a}^\dagger\hat{a}]
$ 
and where $\hat{a}$ and $\hat{a}^\dagger$ are the photon 
annihilation and creation operators, respectively.
In communication theory, this is called a ternary amplitude 
shift keyed (3ASK) signal set\cite{Osa96}. The problem is the 
maximization of the von Neumann entropy $S(\hat{\rho}(\bfxi))$ 
with respect to $\bfxi$. We can assume that a priori 
probability of signal 1 is a certain $\xi_1 = \xi^\prime_1$ at 
first. Under this condition, we take a distribution of fixed 
$\xi_1^\prime$ and arbitrary $\xi_2$ and $\xi_3$, so we have
\begin{equation}
	\bfxi =(\xi^\prime_1,\xi_2,\xi_3) \equiv \bfxi^{(1)},
\end{equation}
and its permutation
\begin{equation}
	\bfxi^{(2)} \equiv (\xi^\prime_1,\xi_3,\xi_2).
\end{equation}
In this case, the next relation is obtained.
\begin{equation}
	\hat{\rho}(\bfxi^{(2)}) 
	= \hat{V}\hat{\rho}(\bfxi^{(1)})\hat{V}^\dagger.
\end{equation}
From the invariance for unitary transformation and the 
concavity of von Neumann entropy, we have
\begin{equation}
	S(\hat{\rho}(\bfxi)) 
	\leq S(\hat{\rho}(\bfxi^\prime)) 
	\equiv S^\prime,
\end{equation}
where
\begin{equation}
	\bfxi^\prime 
	= \frac{1}{2}\bfxi^{(1)} + \frac{1}{2}\bfxi^{(2)} 
	= (\xi^\prime_1, \frac{1-\xi^\prime_1}{2},
			\frac{1-\xi^\prime_1}{2}).
\end{equation}
That is, 
the maximum $S^\prime$ under the condition is achieved 
by $\xi_2=\xi_3=(1-\xi_1^\prime)/2$. So, two a priori 
probabilities of signal 2 and 3 are equal at the optimum 
distribution at least. Then, we can transform the original 
problem to the maximization of the von Neumann entropy of the 
following density operator.
\begin{equation}
	\hat{\rho}^\prime (\xi_1) 
	= \xi_1\hat{\rho}_1 
	+ \frac{1-\xi_1}{2}\hat{\rho}_2 
	+ \frac{1-\xi_1}{2}\hat{\rho}_3.
\end{equation}
As a result, the variable becomes only $\xi_1$. The eigenvalues 
of this operator are 
\begin{eqnarray}
	\lambda_{1}(\xi_1) 
	&=& \frac{1}{2}(1-\kappa^4)(1-\xi_1),
	\\
	\lambda_{2}(\xi_1) 
	&=& \left.\frac{1}{4}\right\{
	(1+\xi_1)+\kappa^4(1-\xi_1)
	\nonumber\\
	& &\quad\left. 
	-2\sqrt{
		\frac{1}{4}
		\left\{
			(1+\xi_1) + \kappa^4(1-\xi_1)
		\right\}^2 
		- 2(1-\kappa^2)^2(1-\xi_1)\xi_1
	}
	\ 
	\right\} ,
	\\
	\lambda_{3}(\xi_1) 
	&=& \left.\frac{1}{4}\right\{
	(1+\xi_1)+\kappa^4(1-\xi_1)
	\nonumber\\
	& &\quad\left. 
	+2\sqrt{
		\frac{1}{4}
		\left\{
			(1+\xi_1)+\kappa^4(1-\xi_1)
		\right\}^2 
		- 2(1-\kappa^2)^2(1-\xi_1)\xi_1
	}
	\ 
	\right\} ,
\end{eqnarray}
where $\kappa = \dotp{0}{\alpha} = \exp[-|\alpha|^2/2]$. 
Using these eigenvalues, 
the maximization problem is written as follows:
\begin{equation}
	C = \max_{0\leq \xi_1 \leq1} \left\{
		-\sum_{i=1}^{3}\lambda_{i}(\xi_1)\log\lambda_{i}(\xi_1)
	\right\}.
\end{equation}
Unfortunately, it is still difficult to solve analytically. So 
we examined numerical calculation, and numerical solution is shown in 
FIG.\ref{fig1}.
 From FIG.\ref{fig1} it turns out that $\xi_1 = 0$ when $|\alpha |^2$ is very small, and $\xi_1$ arises when $|\alpha |^2$ exceeds approximately $0.21$. When $|\alpha |^2$ is quite large, all a priori probabilities converge into $1/3$ respectively.

\begin{figure}[htb]
\begin{center}
\epsfig{file=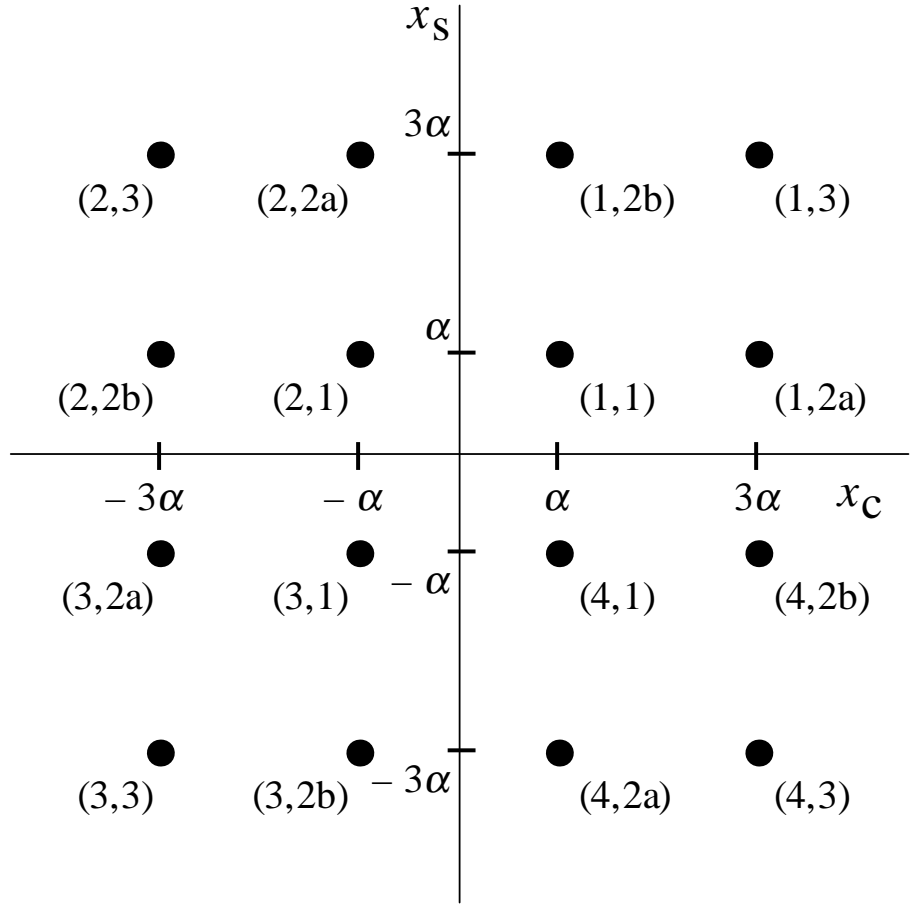,scale=1.5}
\end{center}
\caption{
Signal constellation of 16QAM signal. 
$\hat{x}_{\rm C}\equiv (\hat{a}+\hat{a}^\dagger)/2, 
\hat{x}_{\rm S}\equiv (\hat{a}-\hat{a}^\dagger)/2{\rm i}$. 
Dots stand for amplitudes of coherent states.
}
\label{fig20}
\end{figure}

\begin{figure}[htb]
\begin{center}
\epsfig{file=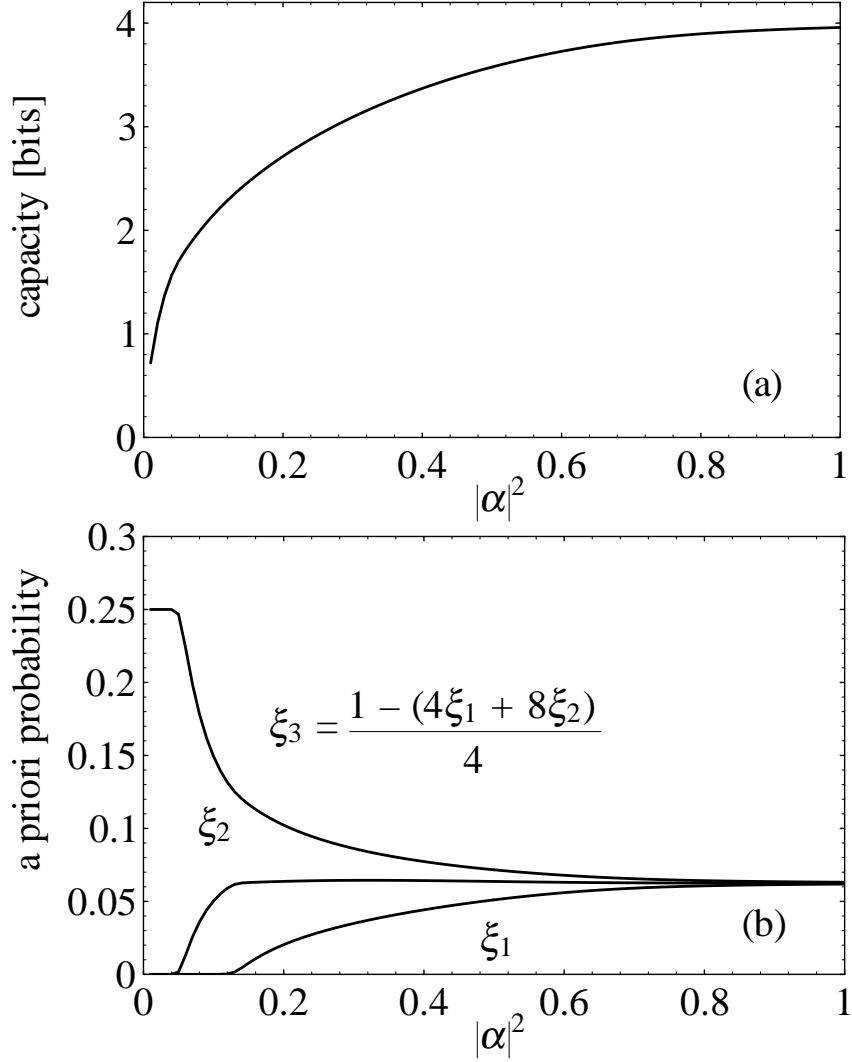,scale=1.5}
\end{center}
\caption{
Channel capacity for 16QAM coherent-state signal.
(a) capacity. (b) optimum distribution
}
\label{fig2}
\end{figure}

As the next example, let us investigate a $16$-ary 
coherent-state 
set called a quadrature amplitude modulated (QAM) signal 
set\cite{Bla}. The complete information source is represented 
as follows:
\begin{eqnarray}
	& &
	\left\{
		\begin{array}{ccccccccc}
			\hat{\rho}_{1,1}, &
			\hat{\rho}_{2,1}, &
			\hat{\rho}_{3,1}, &
			\hat{\rho}_{4,1}, &
			\hat{\rho}_{1,2a}, &
			\hat{\rho}_{2,2a}, &
			\hat{\rho}_{3,2a}, &
			\hat{\rho}_{4,2a}, &
			\hspace{2em}\\
			\xi_{1,1}, &
			\xi_{2,1}, &
			\xi_{3,1}, &
			\xi_{4,1}, &
			\xi_{1,2a}, &
			\xi_{2,2a}, &
			\xi_{3,2a}, &
			\xi_{4,2a}, &
			\hspace{2em}
		\end{array}
	\right.
	\nonumber\\
	& &
	\left.
		\begin{array}{ccccccccc}
			\hspace{1em} &
			\hat{\rho}_{1,2b}, &
			\hat{\rho}_{2,2b}, &
			\hat{\rho}_{3,2b}, &
			\hat{\rho}_{4,2b}, &
			\hat{\rho}_{1,3}, &
			\hat{\rho}_{2,3}, &
			\hat{\rho}_{3,3}, &
			\hat{\rho}_{4,3}
			\\
			\hspace{1em} &
			\xi_{1,2b}, &
			\xi_{2,2b}, &
			\xi_{3,2b}, &
			\xi_{4,2b}, &
			\xi_{1,3}, &
			\xi_{2,3}, &
			\xi_{3,3}, &
			\xi_{4,3}
		\end{array}
	\right\},
\end{eqnarray}
where
\begin{equation}
	\left\{
		\begin{array}{l}
			\begin{array}{ccl}
				\hat{\rho}_{1,1} &=&
\dyad{\alpha + {\rm i}\alpha}{\alpha + {\rm i}\alpha},\\
				\hat{\rho}_{1,2a} &=&
\dyad{3\alpha + {\rm i}\alpha}{3\alpha + {\rm i}\alpha},\\
				\hat{\rho}_{1,2b} &=&
\dyad{\alpha + 3{\rm i}\alpha}{\alpha + 3{\rm i}\alpha},\\
				\hat{\rho}_{1,3} &=&
\dyad{3\alpha + 3{\rm i}\alpha}{3\alpha + 3{\rm i}\alpha},
			\end{array}
			\\
			\begin{array}{cclccl}
				\hat{\rho}_{2,1} &=&
				\hat{V}_A\hat{\rho}_{1,1}\hat{V}_A^{\dagger},&
				\hat{\rho}_{3,1} &=&
				\hat{V}_A^{2}\hat{\rho}_{1,1}\hat{V}_A^{\dagger 2},\\
				\hat{\rho}_{4,1} &=&
				\hat{V}_A^{3}\hat{\rho}_{1,1}\hat{V}_A^{\dagger 3},&
				\hat{\rho}_{2,2a} &=&
				\hat{V}_A\hat{\rho}_{1,2a}\hat{V}_A^{\dagger},\\
				\hat{\rho}_{3,2a} &=&
				\hat{V}_A^{2}\hat{\rho}_{1,2a}\hat{V}_A^{\dagger 2},&
				\hat{\rho}_{4,2a} &=&
				\hat{V}_A^{3}\hat{\rho}_{1,2a}\hat{V}_A^{\dagger 3},\\
				\hat{\rho}_{2,2b} &=&
				\hat{V}_A\hat{\rho}_{1,2b}\hat{V}_A^{\dagger},&
				\hat{\rho}_{3,2b} &=&
				\hat{V}_A^{2}\hat{\rho}_{1,2b}\hat{V}_A^{\dagger 2},\\
				\hat{\rho}_{4,2b} &=&
				\hat{V}_A^{3}\hat{\rho}_{1,2b}\hat{V}_A^{\dagger 3},&
				\hat{\rho}_{2,3} &=&
				\hat{V}_A\hat{\rho}_{1,3}\hat{V}_A^{\dagger},\\
				\hat{\rho}_{3,3} &=&
				\hat{V}_A^{2}\hat{\rho}_{1,3}\hat{V}_A^{\dagger 2},&
				\hat{\rho}_{4,3} &=&
				\hat{V}_A^{3}\hat{\rho}_{1,3}\hat{V}_A^{\dagger 3},
			\end{array}
		\end{array}
	\right.
\end{equation}
and where $\alpha$ is real and 
\begin{equation}
	\hat{V}_A 
	= \exp\left[
		-\frac{\pi{\rm i}}{2}\hat{a}^\dagger\hat{a}
	\right].
\end{equation}
Signal constellation of 16QAM is shown in FIG.\ref{fig20}. 
Based on $\hat{V}_A$ and the concavity of 
von Neumann entropy, we have 
\begin{equation}
	S(\hat{\rho}(\bfxi)) \leq S(\hat{\rho}(\bfxi^\prime)),
	\quad \forall \bfxi,
\end{equation}
where
\begin{eqnarray}
	\bfxi^\prime &=&
	(\xi_1, \xi_1, \xi_1, \xi_1,
	\xi_{2a}, \xi_{2a}, \xi_{2a},\xi_{2a},
	\nonumber\\
	& &
	\quad
	\xi_{2b}, \xi_{2b}, \xi_{2b}, \xi_{2b},
	\xi_3, \xi_3, \xi_3, \xi_3),
\end{eqnarray}
and where
\begin{equation}
	\left\{
		\begin{array}{ccl}
			\xi_1 &\equiv&
			(\xi_{1,1}+\xi_{2,1}+\xi_{3,1}+\xi_{4,1})/4,\\
			\xi_{2a} &\equiv&
			(\xi_{1,2a}+\xi_{2,2a}+\xi_{3,2a}+\xi_{4,2a})/4,\\
			\xi_{2b} &\equiv&
			(\xi_{1,2b}+\xi_{2,2b}+\xi_{3,2b}+\xi_{4,2b})/4,\\
			\xi_3 &\equiv&
			(\xi_{1,3}+\xi_{2,3}+\xi_{3,3}+\xi_{4,3})/4.
		\end{array}
	\right.
\end{equation}
Here we define the anti-unitary operator $\hat{V}_B$ by
\begin{equation}
	\hat{V}_B \ket{\beta} = \ket{\beta^\ast},
\end{equation}
where $\beta^\ast$ means the complex conjugate of $\beta$. 
Because of an invariance of von Neumann entropy for the 
anti-unitary transformation $\hat{V}_{B}$ and the concavity, we 
have the following inequality:
\begin{equation}
	S(\hat{\rho}(\bfxi^\prime)) 
	\leq S(\hat{\rho}(\bfxi^{\prime\prime})),
\end{equation}
where
\begin{eqnarray}
	\bfxi^{\prime\prime} &=&
	(\xi_1, \xi_1, \xi_1, \xi_1, \xi_2, \xi_2, \xi_2, \xi_2,
	\nonumber\\
	& &
	\quad
	\xi_2, \xi_2, \xi_2, \xi_2, \xi_3, \xi_3, \xi_3, \xi_3),
\end{eqnarray}
and where $\xi_2 \equiv (\xi_{2a} + \xi_{2b})/2$. That is, the 
optimization parameters becomes only $\xi_1$ and $\xi_2$, since 
$\xi_3 = (1-4\xi_1-8\xi_2)/4$. Then, the maximization problem 
of the von Neumann entropy for 16QAM coherent-state signal is 
written as
\begin{equation}
	C = \max_{0\leq \xi_1+2\xi_2 
	\leq \frac{1}{4}}S(\hat{\rho}(\bfxi^{\prime\prime})).
\end{equation}
Although we cannot obtain the analytical solution, it is easy 
to find the solution numerically because the number of 
optimization parameters is only two. Numerical solution is 
shown in FIG.\ref{fig2}.
It can be seen from FIG.\ref{fig2} that the optimum a priori 
distribution is classified into three cases. When $|\alpha |^2$ 
is very small, $\xi_1$ and $\xi_2$ are $0$, respectively. 
When 
$|\alpha |^2 
\stackrel{\displaystyle >}{\raisebox{-1ex}{$\sim$}} 
0.04$, $\xi_2$ arises 
and when 
$|\alpha |^2 
\stackrel{\displaystyle >}{\raisebox{-1ex}{$\sim$}} 
0.12$, $\xi_1$ arises. 
When $|\alpha |^2$ is quite large, $\xi_i$ converges into $1/16$.

As concluding remark, we demonstrated that the classical 
capacity for $M$ symmetric states is achieved by the uniform 
distribution of a priori probabilities. 
 Although the analytical result is 
given if and only if the set of quantum states consist of 
symmetric states, our result could apply to the other types of 
the set
of quantum states, in which we can easily find the numerical 
solution. 
Unfortunately we have no method for more general cases, but 
recently it is pointed out that our result is applicable to the 
group covariant case \cite{USU98}.

The authors would like to thank Dr. M.~Ban, Dr. M.~Sasaki, 
and Dr. C.A.~Fuchs for enlightening discussions and useful 
comments. 
They are also grateful to Dr. T.S.~Usuda and 
Mr. S.~Tastuta for their comment.
%

\end{document}